\documentclass[twocolumn,pra,notitlepage,nofootinbib]{revtex4-2}
\usepackage{graphicx}
\usepackage{amsmath,amssymb}%,setspace}
\usepackage{newtxtext}
\usepackage{newtxmath}
\usepackage{xfrac}
\usepackage[nodayofweek,ddmmyy]{datetime}
\usepackage[x11names]{xcolor}
\usepackage[unicode,colorlinks,citecolor=Blue3,linkcolor=Blue3,bookmarks=true]{hyperref}
\usepackage{ulem}

\newcommand{\eg}{{\it e.g.\,}}

\newcommand{\fulld}[2]{\dfrac{d#1}{d#2}}
\newcommand{\partd}[2]{\dfrac{\partial#1}{\partial#2}}
\newcommand{\Tr}{\mathop{\rm Tr}\nolimits}
\newcommand{\bra}[1]{\langle{#1}|}
\newcommand{\ket}[1]{|{#1}\rangle}
\newcommand{\bracket}[2]{\langle{#1}|{#2}\rangle}
\newcommand{\mean}[1]{\langle{#1}\rangle}

\newcommand{\svector}[2]{\begin{pmatrix}#1 \\ #2 \end{pmatrix}}
\newcommand{\smatrix}[4]{\begin{pmatrix}#1 & #2 \\ #3 & #4\end{pmatrix}}
\newcommand{\sh}{\mathop{\rm sinh}\nolimits}
\newcommand{\ch}{\mathop{\rm cosh}\nolimits}

\begin{document}

%\abovedisplayskip=4pt plus 2pt minus 2pt
%\abovedisplayshortskip=2pt plus 1pt minus 1pt
%\belowdisplayskip=4pt plus 2pt minus 2pt
%\belowdisplayshortskip=2pt plus 1pt minus 1pt

\title{Heisenberg limit in phase measurements: the threshold detection approach}

\author{D.~I.~Salykina}%

%\email{koi257@mail.ru}

\affiliation{Faculty of Physics, M.V.Lomonosov Moscow State University, Leninskie Gory 1, Moscow 119991, Russia}

\affiliation{Russian Quantum Center, Skolkovo IC, Bolshoy Bulvar 30, bld.\ 1, Moscow, 121205, Russia}

\author{V.~S.~Liamin}%

\affiliation{Russian Quantum Center, Skolkovo IC, Bolshoy Bulvar 30, bld.\ 1, Moscow, 121205, Russia}

\affiliation{Moscow Institute of Physics and Technology, 141700 Dolgoprudny, Russia}

\author{V.~L.~Gorshenin}

\affiliation{Russian Quantum Center, Skolkovo IC, Bolshoy Bulvar 30, bld.\ 1, Moscow, 121205, Russia}

\affiliation{Moscow Institute of Physics and Technology, 141700 Dolgoprudny, Russia}

\author{B.~N.~Nougmanov}

\affiliation{Russian Quantum Center, Skolkovo IC, Bolshoy Bulvar 30, bld.\ 1, Moscow, 121205, Russia}

\affiliation{Moscow Institute of Physics and Technology, 141700 Dolgoprudny, Russia}

\author{F.~Ya.~Khalili}

\email{farit.khalili@gmail.com}

\affiliation{Russian Quantum Center, Skolkovo IC, Bolshoy Bulvar 30, bld.\ 1, Moscow, 121205, Russia}

%\date{\currenttime\ of \today}

\begin{abstract}

The ultimate precision of phase estimation is limited by the Heisenberg scaling $\Delta\phi_0 = K/N$, where $K\sim1$ is a numerical prefactor and $N$ is the mean number of photons interacting with the phase shifting object(s). However, achieving this fundamental limit often comes at the cost of an extremely narrow high-sensitive range, rendering schemes impractical. We analyze the precision limits of phase measurements in single- and two-arm optical interferometers with input Gaussian states. We consider two detection methods: conventional homodyne measurement and non-Gaussian threshold detection that saturates the quantum Cramér-Rao bound.

We characterize the performance by two complementary metrics: the peak sensitivity $\Delta\phi_0$ and the width $\delta\phi$ of the high-sensitivity range. We demonstrate that Heisenberg scaling is attainable in all configurations considered. However, we reveal that $\delta\phi$ strongly depends on $K$. We derive an approximate analytic expression that describes this trade-off. We show also that the two-arm interferometer with antisymmetrically squeezed inputs exhibits exceptional performance, simultaneously achieving Heisenberg-limited sensitivity and a broad high-sensitivity range $\delta\phi=\pi/2$.

\end{abstract}

\maketitle

%\tableofcontents

\section{Introduction}

Measurement of the phase of light using optical interferometers is one of the key tasks of experimental physics. The sensitivity of the best modern interferometers is very high and to a major extent is limited by quantum fluctuations of the probing light, see \eg the review papers \cite{Demkowicz_PIO_60_345_2015, Andersen_ch35_2019, 22a1SaKh}.

In particular, if the ordinary coherent state of light is used, then the best possible sensitivity corresponds to the Shot Noise Limit (SNL), equal to
\begin{equation}\label{SNL}
  \Delta\phi_{\rm SNL} = \frac{1}{2\sqrt N} \,,
\end{equation}
where $\Delta\phi$ is the mean square phase measurement error and $N$ is the mean number of photons interacting with the phase shifting object(s). Note that due to the presence of the additional reference beam (see Figs.\,\ref{fig:ifo_1}, \ref{fig:ifo_2}), $N$ is not equal to number of photons generated by light source or registered by the detector.

Better sensitivity, for the same value of $N$, can be achieved by using more advanced quantum states. In particular, it was proposed in Ref.\,\cite{Caves1981} to use Gaussian quadrature-squeezed states for this purpose. It was shown in that work that in the case of moderate squeezing, $e^{2r}\ll N$, where $r$ is the logarithmic squeeze factor, the measurement error $\Delta\phi$ can be suppressed by $e^r$, giving the squeezing-enhanced SNL:
\begin{equation}\label{SQZ}
  \Delta\phi_{\rm sqz} = \frac{e^{-r}}{2\sqrt N} \,.
\end{equation}

The most sensitive contemporary optical interferometers, namely the laser gravitational-wave detectors, like LIGO \cite{LSCsite} or VIRGO \cite{VIRGOsite}, use this approach. As a result, they can measure the relative elongations of their multi-kilometer length arms with precision of about $\sim\!10^{-23}\,\text{Hz}^{-1/2}$ \cite{Jia_Science_385_1318_2024}.

In the ultimate case of very strong squeezing, $e^{2r}\sim N$, the sensitivity could reach the so-called Heisenberg limit (HL) \cite{Bondurant_PRD_30_2548_1984, Demkowicz_PIO_60_345_2015, Andersen_ch35_2019}:
\begin{equation}\label{HL}
  \Delta\phi_{\rm HL} \sim \frac{K}{N} \,,
\end{equation}
The same result could also be obtained using more exotic non-Gaussian quantum states of light \cite{Holland_PRL_71_1355_1993, Bollinger_PRA_54_R4649_1996, Demkowicz_NComm_3_1063_2012, Demkowicz_PIO_60_345_2015, Pezze_PRA_91_032103_2015}.

Unfortunately, the non-Gaussian states are notoriously sensitive to the optical losses and are very hard to prepare \cite{Zurek_RMP_75_715_2003}. Moreover, approaches to achieve the Heisenberg scaling often lead to a narrow phase range \cite{17a1MaKhCh}, forcing scientists to look for alternatives, see, \eg, Refs.\,\cite{20a1ShSaFrMiChKh, Jiao_OE_32_46150_2024, Liu_PRL_135_040801_2025}.

Opposite to the well-established limits \eqref{SNL} and \eqref{SQZ}, the HL is still a subject of discussion. Various specific forms of the HL \eqref{HL} and different values of the parameter $K$ can be found in literature.
%The main reason for this is the absence of the unique Hermitian phase operator, canonically conjugated to the photon number operator $\hat{N}$ \cite{Carruthers_RMP_40_411_1968}. Different workarounds that allow us to overcome this issue were proposed, giving different forms of the HL.
In addition, in some works the sensitivity is calculated assuming given mean photon number, while in others --- assuming given total photon number. Reviews of the various approaches to the HL can be found in Refs.\,\cite{Demkowicz_PIO_60_345_2015, Barbiery_PRXQ_3_010202_2022, Gorecki_2304_14370}.

For example, in Refs.\,\cite{Summy_OC_77_75_1990, Bandilla_QO_3_267_1991, Hall_JMO_40_809_1993, Hall_PRA_85_041802_2012}, the following sensitivity limit was obtained for the given mean photon number $N$ and an exotic non-Gaussian state:
\begin{equation}\label{BandillaLimit}
  \Delta\phi_{\rm HL} \approx \frac{1.38}{N} \,.
\end{equation}
At the same time, it is known that the standard linear interferometer and the Gaussian squeezed states allow to reach the sensitivity noticeably exceeding this value, see \eg Ref.\,\cite{17a1MaKhCh}. This discrepancy can be explained by the influence of a priori information on the measurement, see, in particular, Refs.\,\cite{Hall_NJP_14_033040_2012, Gorecky_PRL_124_030501_2020, Gorecki_2304_14370}. Note, indeed, that the bound \eqref{BandillaLimit} holds for any value of $\phi$, while the linear interferometers typically provide the best sensitivity only in some limited range $\delta\phi$ around the optimal point $\phi_0$. Whether this best sensitivity can be reached in practice crucially depends on the a priory information on $\phi$ and the value of the range $\delta\phi$.

In this paper, we explore the sensitivity limits, achievable by the standard optical interferometers using the Gaussian (squeezed coherent) quantum states. For each considered configuration, we optimize the sensitivity assuming a given value of $N$ and calculate two figures of merit. The first one is the minimum $\Delta\phi_0$ of the measurement error $\Delta\phi$, achievable for the value of $\phi$ corresponding to the optimal tuning of the interferometer. It is this parameter that is typically considered in the literature.

The second and, in our opinion, no less important figure of merit, which we analyze here, is $\delta\phi$. Taking into account that typically $\delta\phi\ll1$, and assuming, without loss of generality, that $\phi_0=0$, it can be defined using the ``full width half minimum'' rule as follows:
\begin{equation}\label{delta_phi_def}
  (\Delta\phi)^2
  \approx (\Delta\phi_0)^2\times\biggl[1 + \frac{4\phi^2}{(\delta\phi)^2}\biggr] ,
\end{equation}
where $\phi$ is the actual value of the phase.

We consider two most important from the practical point of view interferometric topologies, the single-arm and the antisymmetric two-arm ones (see \eg the review \cite{22a1SaKh}). The conceptually more simple single-arm topology, see Fig.\,\ref{fig:ifo_1}, directly implements measurement of the phase shift $\phi$ in a harmonic oscillator mode.

In the two-arm configuration, shown in Fig.\,\ref{fig:ifo_2} (the Mach-Zehnder topology is depicted; it is known, however, that the Michelson one is equivalent to it), the phase shifts are introduced antisymmetrically into the first and the second arms, respectively, and both beamsplitters are the balanced 50\%/50\% ones. This configuration is insensitive to the common phase shift and therefore more tolerant to technical noises and drifts. Due to this reason, it is used, in particular, in the GW detectors \cite{12a1DaKh, CQG_32_7_074001_2015}. Here we, in addition to the single-squeezed two-arm interferometer, first proposed in Ref.\,\cite{Caves1981}, consider also the  antisymmetrically double-squeezed version, showing that it provides uniquely broad range of good sensitivity $\delta\phi=\pi/2$.

We consider two types of specific measurement procedures. The first one is the well-known homodyne detector, which is widely used in high-precision experiments. Yet another commonly used detection procedure is the photon number counting, see \eg Refs.\,\cite{Pezze_PRL_100_073601_2008, Demkowicz_PIO_60_345_2015, Sparaciari_PRA_93_023810_2016, Ataman_PRA_98_043856_2018}; we do not consider it here because it gives the results that are close but inferior to the ones of the homodyne detection.

The second procedure which we discuss here is the non-linear threshold detector proposed in Ref.\,\cite{Helstrom_IC_10_254_1967} for detection of binary (yes/no) signals. Here we adapt it to the problem of estimation of a continuous-valued variable (the phase). The importance of this type of measurement stems from the fact that for one given value of $\phi$, it automatically saturates the quantum Cramer-Rao bound (QCRB), which gives the fundamental lower bound for the measurement imprecision \cite{HelstromBook}. At the same time, opposite to the pure QCRB approach, the threshold detector approach allows us to calculate the $\delta\phi$.

This scheme (without using the term ``threshold detector'') was discussed in Ref.\,\cite{Oh_npjQI_5_10_2019}, but only for the single-arm interferometer and for the best sensitivity point $\phi=\phi_0$. Semi-gedanken implementation of this measurement, based on the second order optical nonlinearity, was proposed in Ref.\,\cite{Yanagimoto_Optica_11_896_2024}. Here we extend it to the two-arm configuration. In addition, for both single- and two-arm configurations, following the general logic of this paper, we calculate how the sensitivity degrades if $\phi\ne\phi_0$, providing thus the values of $\delta\phi$.

In order to avoid overloading this paper by the issues that do not relate directly to its main topics, we limit ourselves to the single-mode (in the two-arm interferometer case, two-mode) approximation. This approach is typically used in the literature, see \eg the reviews \cite{Demkowicz_PIO_60_345_2015, Andersen_ch35_2019, 22a1SaKh}. Its adaptation to the continuous-wave case can be found, \eg, in the review \cite{12a1DaKh}. The corresponding spectral forms of QCRB and HL are considered, \eg, in Ref.\,\cite{Loughlin_FiO+LS_2025}. For the same reason, we do not consider the optical losses here. The fundamental quantum interferometry bound for the squeezed light enhanced interferometers can be found in Ref.\,\cite{Demkowicz_PRA_88_041802_2013}.

This paper is organized as follows. In the introductory Sections \ref{sec:CR_TD}  and \ref{sec:ifos} we remind the readers the concepts of QCRB and threshold measurement, as well as the interferometric topologies considered here, and also introduce the main notations used throughout this paper. In Sections \ref{sec:ifo_1} and \ref{sec:ifo_2} we analyze the sensitivity of single-arm and two-arm interferometers, respectively. In Sec.\,\ref{sec:discussion}, we summarize the obtained results and formulate an analytical equation approximately describing the inter-dependence between $\delta\phi$ and the factor $K$.

\section{Quantum Cramer-Rao bound and the threshold detector}\label{sec:CR_TD}

\paragraph{Quantum Cramer-Rao bound.}

Let $\hat{\rho}(\phi)$ be the density operator of an object, depending on the real parameter $\phi$ (\eg the phase) that has to be estimated. It is known, see Sec.\,VIII of the monograph Ref.\,\cite{HelstromBook}, that the lower bound for the variance of an unbiased estimate of $\phi$ exists, which does not depend on the measurement procedure and has the following form:
\begin{equation}\label{CR_gen}
  (\Delta\phi_{\rm QCRB})^2 = \frac{1}{\Tr[\hat{\rho}(\phi)\mathcal{L}^2(\phi)]} \,,
\end{equation}
where $\hat{\mathcal{L}}(\phi)$ is the symmetric logarithmic derivative  defined by the following equation:
\begin{equation}\label{QCRB}
  \partd{\hat{\rho}(\phi)}{\phi} = \hat{\rho}(\phi)\circ\hat{\mathcal{L}}(\phi) \,.
\end{equation}
Here and in the rest of this paper ``$\circ$'' means symmetrized product, that is, for any operators $\hat{Q}$ and $\hat{P}$,
\begin{equation}
  \hat{Q}\circ\hat{P} = \frac{1}{2}(\hat{Q}\hat{P} + \hat{P}\hat{Q}) \,.
\end{equation}

We assume in this paper that the quantum state of the probing light is a pure one and has the following form:
\begin{equation}\label{rho_cov}
  \hat{\rho}(\phi)
  = \hat{\mathcal{R}}(\phi)\ket{\psi_0}\bra{\psi_0}\hat{\mathcal{R}}^\dag(\phi) \,,
\end{equation}
where
\begin{equation}\label{R_of_phi}
  \hat{\mathcal{R}}(\phi) = e^{-i\hat{\mathcal{N}}\phi}
\end{equation}
is the unitary phase shift operator and $\hat{\mathcal{N}}$ is a Hermitian operator. It was shown in Ref.\,\cite{HelstromBook} that in this case Eq.\,\eqref{CR_gen} can be presented in the following closed form:
\begin{equation}\label{CR1}
  (\Delta\phi_{\rm QCRB})^2 = \frac{1}{4(\Delta\mathcal{N})^2} \,,
\end{equation}
where $(\Delta\mathcal{N})^2$ is the variance of $\hat{\mathcal{N}}$ in the state \eqref{rho_cov}.

It is known that the measurement of $\mathcal{L}$ saturates the QCRB, see Refs.\,\cite{HelstromBook, Oh_npjQI_5_10_2019}. Note however that the operator $\hat{\mathcal{L}}$ explicitly depends on $\phi$ (while the QCRB itself does not), creating thus the ``vicious loop''.

\paragraph{Threshold detector.}

Let us consider now an adapted to the measurement of a continuous variable version of threshold detection, initially proposed in \cite{Helstrom_IC_10_254_1967} (see also Sec.\,IV.3 of \cite{HelstromBook}). Suppose that some Hermitian operator $\hat{Y}$ is measured, and we are interested in the sensitivity at some given value of $\phi=0$. Using the error propagation approach, the estimation error $\Delta\phi_0$, corresponding to the vanishingly small values of $\phi\to0$ (the threshold sensitivity) can be calculated as follows:
\begin{equation}\label{d2phi_Y}
  (\Delta\phi_0)^2 = \frac{(\Delta Y)^2}{G^2}\biggr|_{\phi\to0} \,,
\end{equation}
where
\begin{equation}
  (\Delta Y)^2 = \Tr[\hat{\rho}(\phi)\hat{Y}^2] - \bigl(\Tr[\hat{\rho}(\phi)\hat{Y}]\bigr)^2
\end{equation}
is the variance of $\hat{Y}$ and
\begin{gather}\label{G_0}
  G = \partd{\Tr[\hat{\rho}(\phi)\hat{Y}]}{\phi}
\end{gather}
is the gain factor.

It can be shown that the minimum of the uncertainty \eqref{d2phi_Y} is provided by the operator $\hat{Y}$, which satisfies the following equation:
\begin{equation}\label{Helstrom+}
  \lim_{\phi\to0}\partd{\hat{\rho}(\phi)}{\phi} = \hat{\rho}(0)\circ\hat{Y} \,.
\end{equation}
It follows from Eqs.\,\eqref{G_0} and \eqref{Helstrom+} that
\begin{equation}
  \Tr[\rho(0)\hat{Y}] = 0 \,, \quad G\bigr|_{\phi=0} = (\Delta Y)^2\bigr|_{\phi=0} \,
\end{equation}
and therefore,
\begin{equation}\label{TD_short}
  (\Delta\phi_0)^2 = \frac{1}{(\Delta Y)^2}\biggr|_{\phi\to0} \,.
\end{equation}

It is easy to note that up to the notations, in the case of $\phi\to0$, Eq.\,\eqref{QCRB} is identical to Eq.\,\eqref{Helstrom+}. Therefore, the measurement of $\hat{Y}$ saturates the QCRB in this particular case, compare Eqs.\,\eqref{CR_gen} and \eqref{TD_short}.

It is natural to expect that for a smooth $\hat{\rho}(\phi)$ and sufficiently small values of $\phi$, this measurement should provide a good sensitivity, approaching the QCRB, albeit may be not reaching it exactly. The corresponding measurement error in this case can be calculated using the generic error propagation formula
\begin{equation}\label{d2phi_Y_phi}
  (\Delta\phi)^2 = \frac{(\Delta Y)^2}{G^2} \,.
\end{equation}
This approach will be used in the next sections.

\section{Interferometers topologies}\label{sec:ifos}

\paragraph{Single-arm interferometer.}

\begin{figure}
  \includegraphics[scale=0.8]{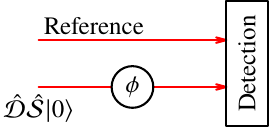}
  \caption{The single-arm interferometer.}\label{fig:ifo_1}
\end{figure}

The scheme of the single-arm interferometer is shown in Fig.\,\ref{fig:ifo_1}. We suppose that the input beam is prepared in the Gaussian squeezed coherent state
\begin{equation}\label{Psi0_1}
  \ket{\psi_0} = \hat{\mathcal{D}}\hat{\mathcal{S}}\ket{0} \,.
\end{equation}
Here $\ket{0}$ is the ground state,
\begin{equation}\label{D1}
  \hat{\mathcal{D}} = e^{\alpha(\hat{a}^\dag - \hat{a})} \,
\end{equation}
is the displacement operator, where we assume without limiting the generality that the displacement parameter $\alpha$ is real,  $\hat{a}$, $\hat{a}^\dag$ are, respectively,  the annihilation and creation operators, and
\begin{equation}\label{S1}
  \hat{\mathcal{S}} = e^{r(\hat{a}^\dag{}^2 - \hat{a}^2)/2} \,
\end{equation}
is the squeeze operator with the real and non-negative squeeze factor $r$. This choice of $r$ corresponds to the squeezed phase quadrature of the input light and therefore to the best phase sensitivity for a given $r$, see Refs.\,\cite{Caves1981, 22a1SaKh}. The mean value $N$ and the variance $(\Delta N)^2$ of the number of quanta
\begin{equation}
  \hat{N}=\hat{a}^\dag\hat{a}
\end{equation}
in the state \eqref{Psi0_1} are equal to
%\begin{subequations}\label{NdN1}
  \begin{gather}
    N = \alpha^2 + \sh^2r \,, \label{mean1}\\
    (\Delta N)^2 = \alpha^2e^{2r} + \frac{1}{2}\sh^22r \,. \label{var1}
  \end{gather}
%\end{subequations}

Introduce the Hermitian amplitude and phase quadrature operators of the input field  $\hat{x}$, $\hat{p}$, defined by
\begin{equation}
  \hat{a} = \frac{\hat{x} + i\hat{p}}{\sqrt{2}} \,,
\end{equation}
and having the variances equal to $1/2$. The input-output relations for these quadratures are the following, see \eg Ref.\,\cite{22a1SaKh}:
\begin{multline}\label{io_1}
  \svector{\hat{x}_{\rm out}}{\hat{p}_{\rm out}}
  = \hat{\mathcal{U}}^\dag(\phi)\svector{\hat{x}}{\hat{p}}\hat{\mathcal{U}}(\phi) \\
  = \svector{(\sqrt{2}\alpha + \hat{x}e^r)\cos\phi + \hat{p}e^{-r}\sin\phi}
      {-(\sqrt{2}\alpha + \hat{x}e^r)\sin\phi + \hat{p}e^{-r}\cos\phi},
\end{multline}
where $\hat{x}_{\rm out}$,  $\hat{p}_{\rm out}$ are the corresponding quadrature operators of the output field,
\begin{equation}
  \hat{\mathcal{U}}(\phi) = \hat{\mathcal{R}}(\phi)\hat{\mathcal{D}}\hat{\mathcal{S}}
\end{equation}
is the evolution operator,
\begin{equation}\label{calR_1}
  \hat{\mathcal{R}}(\phi) = e^{-i\hat{N}\phi}
\end{equation}
is the phase shift operator.

\paragraph{Two-arm interferometer.}

\begin{figure}
  \includegraphics[scale=0.8]{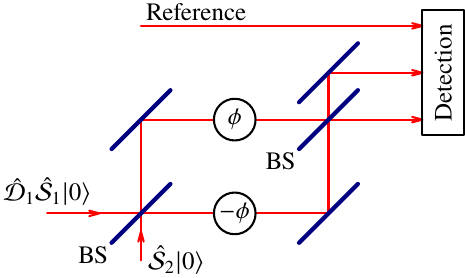}
  \caption{The antisymmetric two-arm interferometer. BS: 50\%:50\% beamsplitters.}\label{fig:ifo_2}
\end{figure}

The antisymmetric two-arm interferometer which we consider here is shown in Fig.\,\ref{fig:ifo_2}. We assume that both beamsplitters are described by the following reflectivity/transmissivity matrix:
\begin{equation}\label{bbB}
  \mathbb{B} = \frac{1}{\sqrt{2}}\smatrix{1}{1}{1}{-1} ,
\end{equation}
and in the absence of the phase shifts $\pm\phi$, the arm lengths are equal to each other. In this case, at $\phi=0$ both input states are reproduced at the respective bright and dark output ports (the dark fringe regime).

Note that in the literature, the signal phase shifts are often denoted as $\pm\phi/2$.  Here we use the ``$\pm\phi$'' convention, because it is more consistent with the single arm case, providing the same values of the SNL \eqref{SNL} and the squeezing-enhanced SNL \eqref{SQZ} as the single-arm interferometer.

Following Refs.\,\cite{Caves1981, Lang_PRL_111_173601_2013}, we assume that the squeezed coherent state is injected into the first (bright) input port, and the squeezed vacuum state --- into the second (dark) input port. This combination gives the equal values of the mean optical power in two arms, keeping the (anti-) symmetry of the scheme intact, see Refs.\, \cite{Hofmann_PRA_79_033822_2009, 22a1SaKh}.

The corresponding two-mode input wave function is the following:
\begin{equation}\label{Psi0_2}
  \ket{\psi_{0,0}} = \hat{\mathcal{D}_1}\hat{\mathcal{S}_1}\hat{\mathcal{S}_2}\ket{0,0} \,.
\end{equation}
Here $\ket{0,0}$ is the two-mode ground state,
\begin{equation}\label{D2}
  \hat{\mathcal{D}}_1 = e^{\alpha(\hat{a}_1^\dag - \hat{a}_1)}
\end{equation}
is the displacement operator of the first (bright) input mode,
\begin{equation}\label{S2}
  \hat{\mathcal{S}}_1 = e^{R(\hat{a}_1^{\dag\,2} - \hat{a}_1^2)/2}\,,\quad
  \hat{\mathcal{S}}_2 = e^{r(\hat{a}_2^{\dag\,2} - \hat{a}_2^2)/2}
\end{equation}
are, respectively, the squeeze operators of the first and second modes, and $\hat{a}_{1,2}$ and $\hat{a}^\dag_{1,2}$ are the corresponding creation and annihilation operators of the input field (before the first beamsplitter). Similarly to the single-arm case, we assume that the parameters $\alpha$, $R$ and $r$ are real (see also Ref.\,\cite{22a1SaKh}). The mean value of the sum photon number in the interferometer is equal to
\begin{equation}\label{mean2}
    N = \alpha^2 + \sh^2R + \sh^2r \,.
\end{equation}

The phase shift operator \eqref{R_of_phi} in the antisymmetric two-arm case has the following form:
\begin{equation}\label{calR_2}
  \hat{\mathcal{R}}_-(\phi) = e^{-i\hat{N}_-\phi} \,,
\end{equation}
where
\begin{equation}\label{N_m}
  \hat{N}_-= \hat{b}_1^\dag\hat{b}_1 - \hat{b}_2^\dag\hat{b}_2
  = \hat{a}_1^\dag\hat{a}_2 + \hat{a}_2^\dag\hat{a}_1 \,,
\end{equation}
is the difference of the photon numbers in the arms and
\begin{equation}
  \hat{b}_1 = \frac{\hat{a}_1 + \hat{a}_2}{\sqrt{2}} \,, \quad
  \hat{b}_2 = \frac{\hat{a}_1 - \hat{a}_2}{\sqrt{2}} \,
\end{equation}
are the annihilation operators for the intracavity fields before the phase shifts, see Eq.\,\eqref{bbB}. It follows from Eqs.\,\eqref{Psi0_2} and \eqref{N_m} that the variance of $\hat{N}_-$ is equal to
\begin{equation}\label{var2}
  (\Delta N_-)^2 = \alpha^2e^{2r} + \sh^2(R+r) \,.
\end{equation}

Similarly to the single-arm case, we introduce the Hermitian amplitude and phase quadrature operators of the input fields  $\hat{x}_{1,2}$, $\hat{p}_{1,2}$, defined by
\begin{equation}
  \hat{a}_{1,2} = \frac{\hat{x}_{1,2} + i\hat{p}_{1,2}}{\sqrt{2}} \,,
\end{equation}
and the corresponding quadrature operators of the output field $\hat{x}_{1,2\,\rm out}$, $\hat{p}_{1,2\,\rm out}$. The input-output relations for these quadratures are the following, see \eg Ref.\,\cite{22a1SaKh}:
\begin{multline}\label{io_2}
  \begin{pmatrix}
    \hat{x}_{1\,\rm out} \\ \hat{p}_{1\,\rm out} \\
      \hat{x}_{2\,\rm out} \\ \hat{p}_{2\,\rm out}
  \end{pmatrix}
  = \hat{\mathcal{U}}_-^\dag(\phi)
        \begin{pmatrix} \hat{x}_1 \\ \hat{p}_1 \\ \hat{x}_2 \\ \hat{p}_2 \end{pmatrix}
      \hat{\mathcal{U}}_-(\phi) \\
  = \begin{pmatrix}
      (\sqrt{2}\alpha + \hat{x}_1e^{R})\cos\phi + \hat{p}_2e^{-r}\sin\phi \\
      -\hat{x}_2e^{r}\sin\phi + \hat{p}_1e^{-R}\cos\phi \\
      \hat{x}_2e^{r}\cos\phi + \hat{p}_1e^{-R}\sin\phi  \\
      -(\sqrt{2}\alpha + \hat{x}_1e^{R})\sin\phi + \hat{p}_2e^{-r}\cos\phi
    \end{pmatrix} ,
\end{multline}
where
\begin{equation}
  \hat{\mathcal{U}}_-(\phi) = \hat{\mathcal{R}}_-(\phi)\hat{\mathcal{D}}_1
    \hat{\mathcal{S}}_1\mathcal{\hat{S}}_2
\end{equation}
is the two-mode evolution operator.

\section{Single-arm interferometer}\label{sec:ifo_1}

\paragraph{Homodyne detection.}

In the case of the small phase shift $|\phi|\ll1$, the phase quadrature of the output beam $\hat{p}_{\rm out}$ carries the major part of the phase information, see Eqs.\,\eqref{io_1}. Therefore, consider the homodyne measurement of this quadrature.  Using the error propagation method, we obtain that the corresponding measurement error is equal to
\begin{equation}\label{dphi1_raw}
  (\Delta\phi)^2 = \frac{(\Delta p_{\rm out})^2}{G^2} \,,
%  = \frac{e^{-2r} + e^{2r}\tan^2\phi}{4\alpha^2} \,,
\end{equation}
where
\begin{equation}\label{d2p_1}
  (\Delta p_{\rm out})^2 = \frac{1}{2}\bigl(e^{-2r}\cos^2\phi + e^{2r}\sin^2\phi\bigr)
\end{equation}
is the variance of $\hat{p}_{\rm out}$ and
\begin{equation}\label{G}
  G = \partd{\mean{\hat{p}_{\rm out}}}{\phi} = -\sqrt{2}\alpha\cos\phi
\end{equation}
is the gain factor, see Eqs.\,\eqref{Psi0_1}, \eqref{io_1}.

If $r\ne0$, then the variance \eqref{d2p_1} depends on $\phi$, and if $r>0$, then it increases with the increase of $\phi$, limiting the high sensitivity range $\delta\phi$. Note also that the two terms in parentheses in Eq.\,\eqref{d2p_1} are related by the uncertainty principle. Therefore, the stronger the squeezing and therefore the better the sensitivity at $\phi=0$, the bigger the second term. As a result, $\delta\phi$ scales with $r$ as $e^{-2r}$.

In order to present Eq.\,\eqref{dphi1_raw} in HL-like form, we minimize it in $\alpha^2$ under the condition \eqref{mean1}. To find the best possible sensitivity, we perform this optimization at $\phi=0$. In this case, the minimum is achieved at
\begin{equation}
  \alpha^2 = \frac{N(N+1)}{2N+1} \,.
\end{equation}
In this case, about half of the optical power is provided by the coherent amplitude and the other part --- by the squeezing, see Eq.\,\eqref{mean1}. It is interesting that the same result was obtained in Ref.\,\cite{Pezze_PRL_100_073601_2008} for the photon number counting case.

Substitution of this value into Eq.\,\eqref{dphi1_raw} gives:
\begin{equation}\label{dphi_HD_1}
  (\Delta\phi)^2 = \frac{1 + (2N+1)^2\tan^2\phi}{4N(N+1)} \,.
\end{equation}

\paragraph{QCRB.}%\label{sec:TD_1}

It follows from Eqs.\,\eqref{CR1}, \eqref{R_of_phi}, and \eqref{calR_1} that the QCRB in the single-arm interferometer case is equal to
\begin{equation}\label{HL_TD_1}
  (\Delta\phi)^2 = \frac{1}{4(\Delta N)^2} \,,
\end{equation}
where the photon number variance is given by Eq.\,\eqref{var1}. Minimizing it in $\alpha^2$ under the condition \eqref{mean1} and taking into account that $\alpha^2\ge0$, we obtain that the minimum is provided by
\begin{equation}\label{TD_1_opt}
  \alpha=0  \,.
\end{equation}
This means that the optimization procedure cancels the information on $\phi$ that could be provided by the mean value of $\hat{p}_{\rm out}$ (see Eq.\,\eqref{G}) and use only the   information provided by the variance \eqref{d2p_1}. As a result, about twice as small value of $(\Delta\phi_0)^2$, in comparison with the homodyne detector, can be achieved:
\begin{equation}\label{Dphi_single}
  (\Delta\phi_0)^2 = \frac{1}{8N(N+1)} \,.
\end{equation}

\paragraph{Threshold detector.}

In order to calculate the measurement error for values of $\phi\ne0$, we derive here the explicit form of the measured observable $\hat{Y}$, see Eq.\,\eqref{Helstrom+}.

In the Schr\"odinger picture, the output field of the interferometer is described by the following density operator:
\begin{equation}\label{rho_1_out}
  \hat{\rho}(\phi) = \mathcal{R}(\phi)\ket{\psi_0}\bra{\psi_0}\mathcal{R}^\dag(\phi) \,,
\end{equation}
where the initial wave function $\ket{\psi_0}$ is given by Eq.\,\eqref{Psi0_1}. By substituting this density operator into Eq.\,\eqref{Helstrom+}, we obtain:
\begin{equation}\label{Helstrom_1}
  i[\hat{\rho}(0),\hat{N}] = \hat{\rho}(0)\circ\hat{Y} \,.
\end{equation}

Consider then a set of quantum states $\{\ket{\psi_k}\}$, with $k=1,2,\dots$, complementing $\ket{\psi_0}$ to the full orthonormal set:
\begin{equation}
  \bracket{\psi_k}{\psi_l} = \delta_{kl} \,,\quad
  \sum_{k=0}^\infty\ket{\psi_k}\bra{\psi_k} = \hat{I} \,,\quad
  k=0,1,\dots
\end{equation}
In this representation, Eq.\,\eqref{Helstrom_1} can be solved explicitly, giving:
\begin{equation}\label{gen_Y_1}
  \bra{\psi_k}\hat{Y}\ket{\psi_l} = \begin{cases}
      0 \,, & k=l=0 \,, \\
      2i\bra{\psi_0}\hat{N}\ket{\psi_l} \,, & k=0\,,\ l>0 \,, \\
      -2i\bra{\psi_k}\hat{N}\ket{\psi_0} \,, & k>0 \,,\ l=0 \,, \\
      \text{undefined} \,, & k>0\,,\ l>0 \,.
    \end{cases}
\end{equation}

In order to build the explicit form of the operator $\hat{Y}$, we have to define the set $\{\ket{\psi_k}\}$. It can be done in different ways. Here, as the natural orthonormal extension of the state \eqref{Psi0_1}, we consider the set of squeezed and displaced Fock states:
\begin{equation}\label{psi_n_1}
  \ket{\psi_n} = \hat{\mathcal{D}}\hat{\mathcal{S}}\ket{n} \,,\quad n = 0,1,\dots,
\end{equation}
where the operators $\hat{\mathcal{D}}$ and $\hat{\mathcal{S}}$ are given by Eqs.\,\eqref{D1} and $\eqref{S1}$, respectively. In this case, at $l>0$,
\begin{equation}
	2i\bra{\psi_0}\hat{N}\ket{\psi_l}
	=	2i\bra{0}\hat{\mathcal{S}}^\dag\hat{\mathcal{D}}^\dag\hat{a}^\dagger\hat{a}
     \hat{\mathcal{D}}\hat{\mathcal{S}}\ket{l}
	= \bra{0}\hat{\mathcal{Y}}\ket{l} \,,
\end{equation}
where
\begin{equation}
  \hat{\mathcal{Y}} = i(2\alpha\hat{a}e^r + \hat{a}^2\sh2r) \,.
\end{equation}
Correspondingly, the form of $\hat{Y}$ that satisfies Eq.\,\eqref{gen_Y_1} for all $k,l\ge0$ is the following:
\begin{multline}\label{Y1_raw}
  \hat{\mathcal{S}}^\dag\hat{\mathcal{D}}^\dag\hat{Y}\hat{\mathcal{D}}\hat{\mathcal{S}}
  = \hat{\mathcal{Y}} + \hat{\mathcal{Y}}^\dag + \hat{a}^\dag\hat{Q}\hat{a} \\
  = -2\hat{p}\circ(\sqrt{2}\alpha e^r + \hat{x}\sh2r) + \hat{a}^\dag\hat{Q}\hat{a}\,,
\end{multline}
where $\hat{Q}$ is an arbitrary Hermitian operator.

The last term in Eq.\,\eqref{Y1_raw} arises due to the non-uniqueness of the operator $\hat{Y}$. The choice of $\hat{Q}$ does not affect the best sensitivity value $\Delta\phi_0$, defined by the QCRB \eqref{CR1}, but could affect the values of $\Delta\phi$ for $\phi\ne0$. In this paper, we consider the case of $\hat{Q}=0$, which corresponds to the simplest bilinear form of $\hat{Y}$. Rolling back the squeeze and displacement operators, we obtain that
\begin{equation}\label{Y_1}
  \hat{Y} = -2\hat{p}\circ(\sqrt{2}\alpha\ch2r + \hat{x}\sh2r) \,.
\end{equation}

In the Heisenberg picture, the value of this operator at the output of the interferometer is equal to
\begin{equation}\label{Y_out_1}
  \hat{Y}_{\rm out} = \hat{\mathcal{U}}^\dag(\phi)\hat{Y}\hat{\mathcal{U}}(\phi)
  = -2\hat{p}_{\rm out}\circ(\sqrt{2}\alpha\ch2r + \hat{x}_{\rm out}\sh2r) ,
\end{equation}
see Eqs.\,\eqref{io_1}.

A slightly refined version of possible scheme for measurement of operator $\hat Y$, proposed in Ref.\,\cite{Yanagimoto_Optica_11_896_2024}, is briefly considered in App.\ref{app:implementation}.

Now we are in a position to calculate the phase measurement error for all values of $\phi$, using the error propagation method. We assume the same value of $\alpha=0$ that gives the best possible sensitivity at $\phi=0$, see Eq.\,\eqref{TD_1_opt}. In this case, it follows from Eqs.\,\eqref{mean1}, \eqref{io_1}, and \eqref{Y_out_1} that the mean value and the variance of $\hat{Y}_{\rm out}$ are equal to
\begin{subequations}
  \begin{gather}
    \mean{\hat{Y}_{\rm out}} = 4N(N+1)\sin2\phi \,,\\
    (\Delta Y_{\rm out})^2 = 8[1 + 4N(N+1)\sin^22\phi]N(N+1) \,.
  \end{gather}
\end{subequations}
Therefore,
\begin{equation}\label{dphi_TD_1}
  (\Delta\phi)^2 = \frac{(\Delta Y_{\rm out})^2}{G^2}
  = \frac{1 + 4N(N+1)\sin^22\phi}{8N(N+1)\cos^22\phi} \,,
\end{equation}
where
\begin{equation}
  G = \partd{\mean{\hat{Y}_{\rm out}}}{\phi} = 8N(N+1)\cos2\phi \,.
\end{equation}

\section{Two-arm interferometer}\label{sec:ifo_2}

\paragraph{Homodyne detector.}%\label{sec:HD_2}

Similarly to the single-arm scheme [compare Eqs.\,\eqref{io_1} and \eqref{io_2}], we consider the homodyne measurement of the quadrature $\hat{p}_{2\,{\rm out}}$ which, in the anti-symmetric regime and in the case of the small phase shift, carries the major part of the phase information, see Eq.\,\eqref{io_2}. In this case,
\begin{equation}\label{dphi2_raw}
  (\Delta\phi)^2 = \frac{(\Delta p_{2\,\rm out})^2}{G^2} \,,
%  = \frac{e^{-2r} + e^{2r}\tan^2\phi}{4\alpha^2} \,,
\end{equation}
where
\begin{equation}\label{d2p_2}
  (\Delta p_{2\,\rm out})^2 = \frac{1}{2}\bigl(e^{-2r}\cos^2\phi + e^{2R}\sin^2\phi\bigr)
\end{equation}
is the variance of the measured quadrature and $G=\partial\mean{\hat{p}_{2\,\rm out}}/\partial\phi$ is the gain factor that again is equal to \eqref{G}.

Note that now, opposite to the single-arm case, two components of $\Delta p_{2\,\rm out}$ originate from two independent input modes and are not bound by the uncertainty relation. Therefore, increase of the squeeze factor $r$  does not affect the second term in Eq.\,\eqref{d2p_2}, proportional to $e^{2R}$. As a result, much broader range $\delta\phi$ could be achieved in the two-arm case.

Let us minimize Eq.\,\eqref{dphi2_raw} in $\alpha^2$ under the condition \eqref{mean2} at the best sensitivity point of $\phi=0$. We consider here two most interesting particular cases: the canonical single-squeezed case of $R=0$ of Ref.\,\cite{Caves1981} and the anti-symmetrically double-squeezed one of $R=-r$. We choose the equal by absolute value squeeze factors, because, on the one hand, typically the stronger is the squeezing, the better is the sensitivity, and, on the other hand, the degree of squeezing in both arms evidently is limited by the same technological limitations. The possible third option is the symmetrically double-squeezed one, $R=r$. However, calculations show that it provides the results similar to the considered in Sec.\,\ref{sec:ifo_1} single-arm case, but slightly inferior to it. Therefore, we do not consider it here.

In the first case,
\begin{equation}\label{d2p_2_1}
  (\Delta p_{2\,\rm out})^2 = \frac{1}{2}\bigl(e^{-2r}\cos^2\phi + \sin^2\phi\bigr) \,,
\end{equation}
the optimum is provided by
\begin{equation}
  \alpha^2 = \frac{N(N+1)}{2N+1}
\end{equation}
and is equal to
\begin{equation}\label{dphi_HD_2_1}
  (\Delta\phi)^2 = \frac{1 + (2N+1)\tan^2\phi}{4N(N+1)} \,,
\end{equation}
see also Ref.\,\cite{17a1MaKhCh}.

In the second case, $\Delta p_{2\,\rm out}$ does not depend on $\phi$:
\begin{equation}
  (\Delta p_{2\,\rm out})^2 = \frac{e^{-2r}}{2} \,. \label{d2p_2_2}
\end{equation}
Correspondingly,
\begin{equation}
  \alpha^2 = \frac{N(N+2)}{2(N+1)}
\end{equation}
and
\begin{equation}\label{dphi_HD_2_2}
  (\Delta\phi)^2 = \frac{1}{2N(N+2)\cos^2\phi} \,.
\end{equation}

\paragraph{QCRB.}

It follows from Eqs.\,\eqref{CR1}, \eqref{R_of_phi} and \eqref{calR_2} that the QCRB in the double-arm interferometer case is equal to
\begin{equation}\label{HL_TD_2}
  (\Delta\phi)^2 = \frac{1}{4(\Delta N_-)^2} \,,
\end{equation}
where the photon number variance is given by Eq.\,\eqref{var2}.

We minimize this value under the condition \eqref{mean2} for the same two particular cases that were considered for the homodyne measurement, namely $R=0$ and $R=-r$. If $R=0$, then the minimum is provided by
\begin{equation}\label{a2_HD_2_1}
  \alpha^2 \approx \frac{N+1/4}{2}
\end{equation}
and is equal to
\begin{equation}
  (\Delta\phi_0)^2 \approx \frac{1}{4N(N+3/2)} \,.
\end{equation}
In the second case, correspondingly,
\begin{equation}
  \alpha^2 = \frac{N(N+2)}{2(N+1)}
\end{equation}
and
\begin{equation}
  (\Delta\phi_0)^2 = \frac{1}{2N(N+2)}  \,.
\end{equation}

\paragraph{Threshold detector.}

Similarly to the single-arm interferometer case, in order to calculate the measurement error for values of $\phi\ne0$, we need the explicit form of the measured observable $\hat{Y}$. It can be calculated in the same way as in the single-arm interferometer case, see Sec.\,\ref{sec:ifo_1}. The corresponding calculations are presented in App.\,\ref{app:Y_2}, see Eq.\,\eqref{Y_2_app}.

In the Heisenberg picture, the value of $\hat{Y}$ at the output of the interferometer is equal to
\begin{multline}\label{Y_out_2}
  \hat{Y}_{\rm out} = 2[
      \sqrt{2}\alpha\hat{p}_{2\,\rm out}e^{r-R}\ch(R+r) \\
      + (
          \hat{x}_{1\,\rm out}\hat{p}_{2\,\rm out}e^{r-R}
          + \hat{x}_{2\,\rm out}\hat{p}_{1\,\rm out}e^{R-r}
        )\sh(R+r)
    ] ,
\end{multline}
see Eq.\,\eqref{io_2}.

Consider again the same two particular cases of $R=0$ and $R=-r$. In the first case, in order to simplify the equations, we assume that $|\phi|\ll1$. The corresponding values of the gain factor $G$ and the variance of $Y_{\rm out}$ are calculated in App.\,\ref{app:Y_2}, see Eqs.\,\eqref{G_dY_2}. Substituting the optimization condition \eqref{a2_HD_2_1} into these equations, we obtain that
\begin{equation}\label{dphi_TD_2_1}
  (\Delta\phi)^2 \approx \frac{1 + \tfrac12(9N + 5/4)\phi^2}{4N(N+3/2)} \,.
\end{equation}

In the second particular case of $R=-r$, the nonlinear term in Eq.\,\eqref{Y_out_2} vanishes and $Y_{\rm out}$ becomes proportional to the quadrature $p_{2\,\rm out}$. This means that the homodyne measurement of this quadrature is now the optimal procedure, described by the phase measurement error \eqref{dphi_HD_2_2}.

\section{Discussion}\label{sec:discussion}

\begin{table}
  \begin{ruledtabular}
  \begin{tabular}{c|cc|cc}
                 & \multicolumn{2}{c|}{HD} & \multicolumn{2}{c}{TD} \\
    \hline
                 & $(\Delta\phi_0)^2$ & $(\delta\phi)^2$ &
                         $(\Delta\phi_0)^2$ & $(\delta\phi)^2$ \\
    \hline
      Single-arm  & $\dfrac{1\vphantom{1^1}}{4N(N+1)}$ & $\dfrac{1}{N(N+1)}$ &
                    $\dfrac{1}{8N(N+1)}$ & $\!\!\dfrac{1}{4N(N+1)}$ \\[2ex]
      Two-arm, $R=0$  & $\!\dfrac{1}{4N(N+1)}$ & $\dfrac{2}{N+\sfrac{1}{2}}$ &
                      $\approx\!\!\dfrac{1}{4N(N+\sfrac{3}{2})}$ & $\!\approx\!\!\dfrac{8}{9N+\sfrac{5}{4}}$
                      \\[2ex]
      Two-arm, $R=-r$ & $\dfrac{1}{2N(N+2)}$ & $\sim1$ & $\dfrac{1}{2N(N+2)}$ & $\Bigl(\dfrac{\pi}{2}\Bigr)^2$ \\[1ex]
  \end{tabular}
  \end{ruledtabular}
  \caption{Optimized sensitivity for the three considered interferometric configurations and for the homodyne detection (HD) and threshold detection (TD) measurement schemes.}\label{tab:1}
\end{table}

The values of $\Delta\phi_0$ and $\delta\phi$ for the configurations considered in this paper [see Eqs.\,\eqref{dphi_HD_1}, \eqref{dphi_TD_1}, \eqref{dphi_HD_2_1}, \eqref{dphi_HD_2_2}, \eqref{dphi_TD_2_1}] are presented in Table\,\ref{tab:1}. It can be seen that in all cases, the HL scaling \eqref{HL} can be achieved, but with different values of the factor $K$. It can be seen also that there is an interdependence between the values of $\Delta\phi_0$ and $\delta\phi$:  in all cases except for the single-arm interferometer with the homodyne detection, it can be roughly estimated as follows:
\begin{equation}\label{loglog}
  \log_N(\delta\phi) \sim \log_2(N\Delta\phi_0) + \frac{1}{2} \,.
\end{equation}
This formula means that a two-fold gain in sensitivity leads to a narrowing of the high sensitivity range by a factor of $\sim\sqrt{N}$.

The single-arm interferometer is characterized by the narrowest range of high sensitivity, $\delta\phi\sim1/N$ for both the homodyne and threshold detection. Note that it is close to $\Delta\phi_0$, that is, in this case a priori knowledge about the phase distribution before the measurement should be as good as measurement precision. This feature could make the use of this regime problematic.

These two features stem from the same origin, namely the enhanced by the squeezing amplitude quadrature of the input light $\hat{x}e^r$, which appears in the measured phase quadrature of the output light $\hat{p}_{\rm out}$ at $\phi\ne0$, see Eq.\,\eqref{io_1}. Note also that in the single-arm interferometer, this term is bound by the uncertainty relation with the term $\hat{p}e^{-r}$ which defines the measurement error at $\phi=0$. As a result, the variance $\Delta p_{\rm out}$ sharply increases with the increase of $\phi$, limiting the value of $\delta\phi$.

At the same time, the dependence of $\Delta p_{\rm out}$ on $\phi$ can be exploited as the additional source of information using more advanced data processing or more sophisticated measurements. In particular, in Ref.\,\cite{Pezze_PRL_100_073601_2008} the use of the maximum likelihood method was considered. The disadvantage of this approach is that it requires multiple measurements, which leads to the additional factor $\sqrt{p}$ in the numerator of the HL \eqref{HL}, where $p$ is the repetitions number (assuming that $N$ is still the total photon number used in all repetitions). Another approach is the use of a non-linear measurement, in particular, of the considered here threshold detector. It is interesting that in this case the optimization procedure prefers the variance information to the linear one, setting the value of $\alpha$ equal to zero, see Eq.\,\eqref{TD_1_opt}.

In the two-arm interferometer case, the measured output quadrature $\hat{p}_{\rm out}$ depends on the quadratures $\hat{x}_1$ and $\hat{p}_2$ of the two independently prepared input modes that are not bound by the uncertainty relation. As a result, the squeezing of $\hat{p}_2$ does not affect uncertainty of $\hat{x}_1$ and therefore does not increase the value of $\Delta p_{2\,{\rm out}}$ at $\phi\ne0$. In particular, if the first mode is not squeezed, $R=0$, then the value of $\delta\phi$ scales only as $1/\sqrt{N}$, that is, in this case, $\delta\phi\gg\Delta\phi_0$. This feature allows to use, before the main measurement, a preliminary ``ranging'' shot noise limited one (without squeezing in both arms), which is characterized by $\delta\phi\sim1$, thus providing, in the series of two measurement, both broad $\delta\phi$ and the HL sensitivity.

On the other hand, this weaker dependence of the $\Delta p_{2\,{\rm out}}$ on $\phi$ means that less information is available to the nonlinear measurement. Indeed, it can be seen from Table\,\ref{tab:1} that if $R=0$, then the threshold detector provides only the marginally better sensitivity than the homodyne one.

Finally, in the double-squeezed case of $R=-r$, the variance of $\hat{p}_{\rm 2\,out}$ does not depend on $\phi$ at all, see Eq.\,\eqref{d2p_2_2}. As a result, the non-linear term in $\hat{Y}_{\rm out}$ vanishes and the threshold detection scheme reduces to the ordinary homodyne measurement. The high sensitivity range in this case is uniquely broad, $\delta\phi=\pi/2$ (taking into account that in this case the condition $\delta\phi\ll1$ is not valid, we directly use Eq.\,\eqref{dphi_HD_2_2} instead of \eqref{delta_phi_def}).
The price for this is the value of $(\Delta\phi_0)^2$ that is twice as big as in the previous case. However, it is interesting to note that the corresponding HL prefactor $1/\sqrt{2}$ is still smaller than $\approx\!\!1.38$ of the limit \eqref{BandillaLimit}, despite the exotic non-Gaussian nature of the quantum states providing that limit.

\acknowledgments

This work of D.S., V.G., B.N., and F.K. was supported by the Theoretical Physics and Mathematics Advancement Foundation ``BASIS'' Grant \#23-1-1-39-1. Authors thank L.\,Pezze for his valuable remarks on this work.

\appendix

\section{Threshold detection, two-arm interferometer case}\label{app:Y_2}

In the two-arm interferometer case, Eqs.\,\eqref{rho_1_out} and \eqref{Helstrom_1} take the following form:
\begin{gather}
  \hat{\rho}(\phi)
    = \hat{\mathcal{R}}_-(\phi)\hat{\rho}_{0,0}\hat{\mathcal{R}}_-(\phi) \,,
      \label{rho_2_out} \\
  i[\hat{\rho}_{0,0}\hat{N}_-] = \hat{\rho}_{0,0}\circ\hat{Y} \,, \label{Helstrom_2}
\end{gather}
where
\begin{equation}
  \hat{\rho}_{0,0} = \ket{\psi_{0,0}}\bra{\psi_{0,0}}
\end{equation}
Let the states $\ket{\psi_{k_1,k_2}}$, where $k_{1,2}=1,2,\dots$, complement $\ket{\psi_{0,0}}$ to the two-dimensional full orthonormal set:
\begin{equation}
    \bracket{\psi_{k_1,k_2}}{\psi_{l_1,l_2}} = \delta_{k_1,l_1}\delta_{k_2,l_2} \,, \quad
    \sum_{k_1,k_2=0}^\infty\ket{\psi_{k_1,k_2}}\bra{\psi_{k_1,k_2}} = \hat{I} \,.
\end{equation}
Using this representation, Eq.\,\eqref{Helstrom_2} can be solved explicitly:
\begin{multline}\label{gen_Y_2}
  \bra{\psi_{k_1,k_2}}\hat{Y}\ket{\psi_{l_1,l_2}} \\
  = \begin{cases}
      0 \,, & k_1=k_2=l_1=l_2=0 \,, \\
      2i\bra{\psi_{0,0}}\hat{N}_-\ket{\psi_{l_1,l_2}} \,, & k_1=k_2=0\,,\ l_1+l_2>0 \,, \\
      -2i\bra{\psi_{k_1,k_2}}\hat{N}_-\ket{\psi_{0,0}} \,, & k_1+k_2>0 \,,\ l_1=l_2=0 \,, \\
      \text{undefined} \,, & k_1+k_2>0\,,\ l_1+l_2>0 \,.
    \end{cases}
\end{multline}
It follows from this solution that
\begin{equation}
  (\Delta Y)^2 = 4(\Delta N_-)^2 \,.
\end{equation}
As a result, we obtain Eq.\,\eqref{HL_TD_2}.

Let us construct now a possible explicit form of $\hat{Y}$. The two-dimensional analog of the states \eqref{psi_n_1} is the following:
\begin{equation}
  \ket{\psi_{n_1,n_2}}= \hat{\mathcal{V}}\ket{n_1,n_2}\,,\quad n_{1,2} = 0,1,\dots,
\end{equation}
where
\begin{equation}
  \hat{\mathcal{V}} = \hat{\mathcal{D}}_1\hat{\mathcal{S}}_1\mathcal{\hat{S}}_2 \,,
\end{equation}
the operators $\hat{\mathcal{D}}_1$, $\hat{\mathcal{S}}_{1,2}$ are given by Eqs.\,\eqref{D2} and $\eqref{S2}$,
and $\ket{n_1,n_2}$ are the two-dimensional Fock states. Correspondingly, Eq.\,\eqref{gen_Y_2} gives:
\begin{multline}
  \bra{k_1,k_2}\hat{\mathcal{V}}^\dag\hat{Y}\hat{\mathcal{V}}\ket{l_1,l_2} \\
  = \begin{cases}
      0 \,, & k_1=k_2=l_1=l_2 = 0 \,, \\
      \bra{0,0}\hat{\mathcal{Y}}_2\ket{l_1,l_2} \,, &  k_1=k_2=0\,,\ l_1+l_2>0 \,, \\
      \bra{k_1,k_2}\hat{\mathcal{Y}}_2^\dag\ket{0,0} \,, & k_1+k_2>0 \,,\ l_1=l_2=0 \,,\\
      \text{undefined} \,, & k_1+k_2>0\,,\ l_1+l_2>0 \,,
    \end{cases} %\label{Y_ifo2}
\end{multline}
where
\begin{gather}
  \hat{\mathcal{Y}}_2 = 2i(\alpha\hat{a}_2e^r + \hat{a}_1\hat{a}_2\sh R_+) \,, \\
  R_+ = R+r \,.
\end{gather}

We define $\hat{Y}$ for all $k_{1,2}>0, l_{1,2}>0$ as follows:
\begin{equation}\label{Y2_raw}
  \hat{\mathcal{V}}^\dag\hat{Y}\hat{\mathcal{V}}
  = \hat{\mathcal{Y}}_2 + \hat{\mathcal{Y}}_2^\dag
  = -2[\sqrt{2}\alpha\hat{p}_2e^r + (\hat{x}_1\hat{p}_2 + \hat{x}_2\hat{p}_1)\sh R_+]
\end{equation}
(we omit for brevity the additional operator $Q$, see discussion after Eq.\,\eqref{Y1_raw}).
Therefore,
\begin{multline}\label{Y_2_app}
  \hat{Y}  = \hat{\mathcal{V}}(\hat{\mathcal{Y}}_2 + \hat{\mathcal{Y}}_2^\dag)
    \hat{\mathcal{V}}^\dag \\
  = -2[
        \sqrt{2}\alpha\hat{p}_2e^{r-R}\ch(R+r) \\
        + (\hat{x}_1\hat{p}_2e^{r-R} + \hat{x}_2\hat{p}_1e^{R-r})\sh(R+r)
      ] .
\end{multline}

Let now $R=0$ and $|\phi|\ll1$. In this case, the cumbersome but straightforward calculations give that
\begin{subequations}\label{G_dY_2}
  \begin{equation}
    G = \partd{\mean{\hat{Y}_{\rm out}}}{\phi} = 4(\alpha^2e^{2r} + \sh^2r) \,,
  \end{equation}
  \begin{multline}
    (\Delta\hat{Y}_{\rm out})^2 = 4\bigl\{
        \alpha^2e^{2r} + \sh^2r \\
        + \bigl[\alpha^2e^{2r}(\ch r + 2\sh r)^2 + 4\ch^2r\sh^2r\bigr]\phi^2
      \bigr\} .
  \end{multline}
\end{subequations}

\section{Implementation of the threshold detector}\label{app:implementation}

Consider measurement of the following operator:
\begin{equation}
	\hat{Y} = A\hat{p} + B\hat{x}\circ\hat{p} \,,
\end{equation}
where $A, B$ are c-numbers, see Eq.\,\eqref{Y_out_1}. Using the displacement operation:
\begin{equation}
  \mathcal{D}^\dag\hat{x}\mathcal{D} = \hat{x} - A/B \,,
\end{equation}
which can be implemented using a linear beamsplitter, it can be reduced to
\begin{equation}
	\hat{Y}_{\rm red} = B\hat{x}\circ\hat{p}  \,.
\end{equation}

Consider now the second (probe) mode
\begin{equation}
  \hat{b} = \frac{\hat{X} + i\hat{P}}{\sqrt{2}} \,,
\end{equation}
coupled with first (signal) one by the degenerate parametric Hamiltonian equal to
\begin{equation}
	H = \frac{\hbar g}{\sqrt{2}}(\hat{a}^2\hat{b}^\dag + \hat{a}^{\dag 2}\hat{b})
  = \frac{\hbar g}{2}[(\hat{x}^2-\hat{p}^2)\hat{X} + 2(\hat{x}\circ\hat{p})\hat{P}] \,,
\end{equation}
where $g$ is the coupling factor.

The Heisenberg equations of motion for the operators $\hat{X}$ and $\hat{x}\circ\hat{p}$ are the following:
\begin{subequations}
  \begin{gather}
    \fulld{\hat{X}}{t} = g\,\hat{x}\circ\hat{p} \,, \\
    \fulld{(\hat{x}\circ\hat{p})}{t} = -g\,(\hat{x}^2 + \hat{p}^2)\hat{X} \,,
  \end{gather}
\end{subequations}
The second-order iterative solution to these equations can be presented as follows:
\begin{equation}
  \hat{X}(t) = gt(\hat{x}\circ\hat{p}) + \delta\hat{X} \,,
\end{equation}
where
\begin{equation}
  \delta\hat{X} = \biggl[1 - \frac{(gt)^2}{2}(\hat{x}^2 + \hat{p}^2)\biggr]\hat{X}\,.
\end{equation}
Therefore, the measurement error for $\hat{x}\circ\hat{p}$ is equal to
\begin{multline}
  \Delta_{\rm meas}^2 = \frac{\mean{(\delta\hat{X})^2}}{(gt)^2} \\
  = \biggl[
      \frac{1}{(gt)^2} - \mean{\hat{x}^2 + \hat{p}^2}
      + \frac{(gt)^2}{4}\mean{(\hat{x}^2 + \hat{p}^2)^2}
    \biggr]\mean{\hat{X}^2} \,.
\end{multline}
(we assumed here for simplicity that $\mean{\hat{X}}=0$).

Within this second-order approximation, it is proportional to $\mean{\hat{X}^2}$. In Ref.\,\cite{Yanagimoto_Optica_11_896_2024}, it was proposed to use the preliminary squeezing of $\hat{X}$ to reduce $\Delta_{\rm meas}^2$. In addition, it was proposed in that work to use the proportional anti-squeezing of the output value of $\hat{X}$. It follows from our consideration, that this is not necessary. However, the output anti-squeezing, not necessarily complementary to the input squeezing, makes the setup more tolerant to the optical losses in the subsequent stages of the setup, see Ref.\,\cite{17a1MaKhCh}.

The adaptation of this approach to the two mode case can be done in a similar manner if one considers a non-degenerate parametric process.

%\bibliography{khalili_u,biblio_us,abbots_my,LIGO,mqm,qoptics,misc_u}


\end{document}